\documentclass{ws-procs975x65}            
\begin{document}
\title{Detection Prospects of the Cosmic Neutrino Background}

\author{Yu-Feng Li}

\address{Institute of High Energy Physics, Chinese Academy of Sciences, \\
P.O. Box 918, Beijing 100049, China\\
Email: liyufeng@ihep.ac.cn}

\begin{abstract}
The existence of the cosmic neutrino background (C$\nu$B) is a fundamental prediction of the standard Big Bang cosmology.
Although current cosmological probes provide indirect observational evidence, the direct detection of the C$\nu$B
in a laboratory experiment is a great challenge to the present experimental techniques.
We discuss the future prospects for the direct
detection of the C$\nu$B, with the emphasis on the method of captures on beta-decaying nuclei and the PTOLEMY project.
Other possibilities using the electron-capture (EC) decaying nuclei, the annihilation
of extremely high-energy cosmic neutrinos (EHEC$\nu$s) at the $Z$-resonance, and the atomic de-excitation method are also discussed in this review.
\end{abstract}


\bodymatter

\section{Introduction}
\label{intro}
As weakly-interacting and rather stable particles, relic neutrinos were
decoupled from radiation and matter at a temperature of about one MeV and an age of one
second after the Big Bang \cite{XZ}. It is quite similar to the cosmic microwave background (CMB) radiation,
whose formation was at a time of around $3.8\times10^{5}$ years after the Big Bang.
This cosmic neutrino background (C$\nu$B) played an important role in the evolution of the Universe,
and its existence has been indirectly proved from current cosmological data on the
Big Bang nucleosynthesis (BBN), large-scale structures of the cosmos and CMB anisotropies \cite{PDG}.

The properties of the C$\nu$B are tightly related to the properties of the CMB. In particular,
in the absence of lepton asymmetries the temperature and average number density of relic neutrinos
can be expressed as \cite{XZ}
\begin{equation}
T^{}_\nu = \left(\frac{4}{11}\right)^{1/3}T^{}_\gamma\approx 1.945 \; {\rm K}\,,\quad\quad\quad
n^{}_\nu = \frac{9}{11}n^{}_\gamma\approx 336\;{\rm cm}^{-3}\,.
\end{equation}
As a consequence, one predicts the average three-momentum today for each species of the relic neutrino is
very small \cite{Ringwald}:
\begin{equation}
\langle p^{}_\nu \rangle = 3T^{}_\nu \approx 5.8 \; {\rm K}\approx 5 \times 10^{-4}
 \; {\rm eV}\,,
\end{equation}
implying that at least two mass eigenstates of relic neutrinos are already non-relativistic,
no matter whether the neutrino mass spectrum is the normal or inverted hierarchy.

Although cosmological observations provide the indirect evidence for the existence of the C$\nu$B,
direct detection in a laboratory experiment is a great challenge to the present experimental techniques.
Among several possibilities \cite{Ringwald}, the most promising one seems to be the neutrino capture experiment
using radioactive $\beta$-decaying nuclei \cite{Weinberg,Irvine,Cocco,Vogel,Blennow,Kaboth,LLX,LX11,Liao,Long14}. The proposed
PTOLEMY project \cite{PTOLEMY} aims to obtain the sensitivity required to detect the C$\nu$B using 100 grams
of $^3$H as the capture target. Other interesting possibilities include the electron-capture (EC) decaying
nuclei \cite{Cocco2,Lusignoli,LX11EC,LXJCAP}, the annihilation of extremely high-energy cosmic neutrinos (EHEC$\nu$s)
at the $Z$-resonance \cite{Weiler82,Eberle04,Barenboim04}, and the atomic de-excitation method \cite{atomic}.

The remaining parts of this work are organized as follows. In Sec.~2 we introduce the method of captures on the
beta-decaying nuclei and calculate the $\beta$-decay energy spectrum and the relic neutrino capture rate. Sec.~3 is
devoted to flavor effects of the relic neutrino capture spectrum. We shall present a brief description on other
interesting possibilities of the C$\nu$B detection in Sec.~4, and then conclude in Sec.~5.

\section{Captures on the Beta-decaying Nuclei}
\label{detection}

In the presence of $3+n$ species of active and sterile
neutrinos, the flavor eigenstates of three active neutrinos and $n$ sterile neutrinos can be
written as \cite{XZ,PDG}
\begin{eqnarray}
\left(\begin{matrix} \nu^{}_e \cr \nu^{}_\mu \cr \nu^{}_\tau \cr \vdots \cr
\end{matrix}
\right) = \left(\begin{matrix} U^{}_{e1} & U^{}_{e2} & U^{}_{e3} & \cdots \cr
U^{}_{\mu 1} & U^{}_{\mu 2} & U^{}_{\mu 3} & \cdots \cr
U^{}_{\tau 1} & U^{}_{\tau 2} & U^{}_{\tau 3} & \cdots \cr
\vdots & \vdots & \vdots & \ddots \cr \end{matrix} \right)
\left(\begin{matrix} \nu^{}_1 \cr \nu^{}_2 \cr \nu^{}_3 \cr \vdots \cr
\end{matrix}
\right) \; ,
\end{eqnarray}
where $\nu^{}_i$ is a mass eigenstate of active (for $1 \leq i \leq 3$) or sterile (for $4 \leq
i \leq 3 + n$) neutrinos, and $U^{}_{\alpha i}$ stands for an element of the $(3+ n) \times (3+ n)$ neutrino mixing matrix.

\begin{figure}[t]
\begin{center}
\begin{tabular}{c}
\includegraphics*[width=0.6\textwidth]{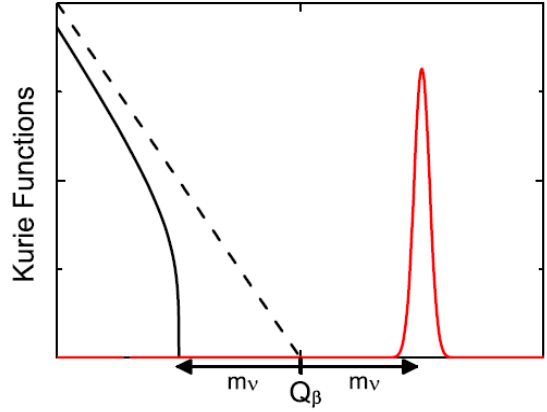}
\end{tabular}
\end{center}
\vspace{-0.5cm}
\caption{Idealized electron spectra for the tritium beta decay and relic neutrino capture. The
dashed and black-solid lines are shown for $\beta$-decay spectra of the massless and massive neutrinos respectively.
The red-solid line with the sharp peak is for the relic neutrino signal.}
\label{capture}
\end{figure}

One of the most promising method to measure the absolute electron neutrino mass is the nuclear $\beta$-decay
\begin{eqnarray}
{\cal N}(A,Z) \to {\cal  N}^\prime (A, Z+1) + e^- + \overline{\nu}^{}_e\,,
\end{eqnarray}
where $A$ and $Z$ are the mass and atomic numbers of the parent nucleus, respectively.
The differential decay rate of a $\beta$-decay can be written as \cite{Weinheimer}
\begin{eqnarray}
\frac{{\rm d} {\lambda}^{}_\beta}{{\rm d}T^{}_e} && =
\int_0^{Q^{}_{\beta}- {\rm min}(m^{}_i)} {\rm d} T^\prime_e \,
\left\{\frac{G^2_{\rm F} \, \cos^2\theta^{}_{\rm C}}{2\pi^3} \,
F\left(Z, E^{}_{e}\right) \, |{\cal M}|^2 E^{}_{e}\sqrt{E^2_e -
m^2_e}   \,  \right .
\nonumber \\
& & \left . \times\left(Q^{}_{\beta} - T^\prime_e\right)\sum^4_{i=1}
\left[ |U^{}_{ei}|^2\sqrt{\left(Q^{}_{\beta}- T^\prime_e \right)^2 -
m_i^2} ~ \Theta\left(Q^{}_{\beta} - T^\prime_e - m^{}_i\right)
\right]\right\}
\nonumber \\
& & \times R\left(T^{}_e, T^\prime_e\right) \; ,
\end{eqnarray}
where $T^\prime_e = E^{}_e - m^{}_e$ denotes the intrinsic kinetic energy
of the outgoing electron, $F(Z, E^{}_{e})$ is the Fermi function,
$|{\cal M}|^2$ is the dimensionless nuclear matrix elements \cite{Weinheimer},
and $\theta^{}_{\rm C} \simeq 13^\circ$ is the Cabibbo angle.
Note that a Gaussian energy resolution function,
\begin{equation}
R(T^{}_{e}, \, T^{\prime}_{e}) = \frac{1}{\sqrt{2\pi} \,\sigma}
\exp\left[-\frac{(T^{}_{e} - T^{\prime}_{e})^2}{2\sigma^2} \right] \; ,
\end{equation}
is implemented in Eq.~(5) to include the finite energy resolution, and the theta function
is adopted to ensure the kinematic requirement. The spectral shape near the $\beta$-decay
endpoint represents a kinetic measurement of the absolute neutrino masses, which can
be understood by comparing the dashed and black solid lines of Fig.~\ref{capture}. The gap
between end points of two black lines stands for a measurement of the effective
electron neutrino mass.

On the other hand, the threshold-less neutrino capture process,
\begin{eqnarray}
{\nu}^{}_e + {\cal N}(A,Z) \to {\cal  N}^\prime (A, Z+1) + e^-\,,
\end{eqnarray}
is located well beyond the end point of the $\beta$-decay, where the signal is characterized by
the monoenergetic kinetic energy of the electron for each neutrino mass
eigenstate. A measurement of the distance between the decay and capture processes will directly
probe the C$\nu$B and constrain or determine the masses and mixing angles.
The differential neutrino capture rate of this process reads
\begin{equation}
\frac{{\rm d} \lambda^{}_{\nu}}{{\rm d} T^{}_{e}} = \sum_i
|U^{}_{ei}|^2 \sigma^{}_{\nu^{}_i} v^{}_{\nu^{}_i}
n^{}_{\nu^{}_i}\,R(T^{}_{e}, \, T^{\prime i}_{e}) \; ,
\end{equation}
where the sum is for all the neutrino mass eigenstates and
\begin{equation}
n^{}_{\nu^{}_i}=\frac{n^{}_{\nu^{}_i}}{\langle n^{}_{\nu^{}_i}\rangle}\cdot\langle n^{}_{\nu^{}_i}\rangle
\equiv\zeta^{}_i \cdot\langle n^{}_{\nu^{}_i}\rangle\,,
\end{equation}
denotes the number density of the relic neutrinos $\nu^{}_i$ around the Earth. The standard Big Bang cosmology gives the
prediction $\langle n^{}_{\nu^{}_i} \rangle \approx \langle n^{}_{\overline{\nu}^{}_i} \rangle \approx 56 ~{\rm cm}^{-3}$
for each species of active neutrinos, and the prediction is also expected to hold for each species
of sterile neutrinos if they could be fully thermalized in the early Universe.
The number density of relic neutrinos around the Earth may be enhanced by the
gravitational clustering effect (i.e., the factor $\zeta^{}_i$) when the neutrino mass is larger than 0.1 eV \cite{Wong}.
In Eq.~(7) the capture cross-section times neutrino velocity can be written as
\begin{eqnarray}
\sigma_{\nu_i} v_{\nu_i}=\frac{2\pi^{2}}{\emph{A}}\cdot\frac{\ln2
}{T_{1/2}}\,,
\end{eqnarray}
where $\emph{A}$ is the nuclear factor characterized by $Q_{\beta}$ and $Z$, and
$T_{1/2}$ is the half-life of the parent nucleus.

Considering the running time and target mass of a particular experiment, the distributions of the numbers of
capture signal and $\beta$-decay background events are expressed, respectively, as
\begin{eqnarray}
\frac{{\rm d} N^{}_{\rm S}}{{\rm d} T^{}_{e}} \hspace{-0.0cm} & = &
\hspace{-0.0cm} {1 \over \lambda^{}_{\beta}} \cdot {{\rm d}
\lambda^{}_{\nu} \over {\rm d} T^{}_{e}} \cdot {\ln 2 \over T^{}_{1/2}} \,
{\bar N}^{}_{\rm T} \, t \; ,
\nonumber \\
\frac{{\rm d} N^{}_{\rm B}}{{\rm d} T^{}_{e}} \hspace{-0.0cm} & = &
\hspace{-0.0cm} {1 \over \lambda^{}_{\beta}} \cdot {{\rm d}
\lambda^{}_{\beta}\over {\rm d} T^{}_{e}} \cdot {\ln 2 \over T^{}_{1/2}} \,
{\bar N}^{}_{\rm T} \, t \; ,
\end{eqnarray}
where ${\bar N}^{}_{\rm T}$ is the averaged number of target
atoms for a given exposure time $t^{}_{}$. Therefore, ${\bar N}^{}_{\rm T}\,t^{}_{}$ gives
the total target factor in the experiment:
\begin{eqnarray}
{\bar N}_{\rm T}\,t^{}_{}=N(0)\cdot\frac{T_{1/2}}{\ln2
}\cdot\left(1-e^{-t\cdot\frac{\ln2}{T_{1/2}}}\right)\,,
\end{eqnarray}
with $N(0)$ being the initial target number at $t=0$.

To get a better signal-to-background ratio, one can investigate different kinds of
candidate nuclei by considering factors of the cross-section, half-life, $\beta$-decay rate,
and the detector energy resolution.
The target nuclei should have the half-life $T_{1/2}$ longer than duration of the exposure time, have the
maximal possible cross-section times neutrino velocity, and have the minimal possible background rate.
An exhaustive survey was done several years ago \cite{Cocco}, and a summary of several candidates
of the $\beta^{-}$-decaying nuclei is shown in Tab.~1, from which one can find $^{3}$H,
$^{106}$Ru, and $^{187}$Re can be possible promising nuclei.
\begin{table}
\tbl{Several candidates of the $\beta^{-}$-decaying nuclei \cite{Cocco}.}
{\begin{tabular}{@{}ccccc@{}}
\toprule
Isotope & $Q_{\beta}$ & Decay type & Half-life & $\sigma_{\nu_i}\cdot v_{\nu_i}$ \\
& (keV) & & (sec) & ($10^{-41}$ cm$^2$) \\
\colrule
$^{3}$H     &  18.591 & $\beta^-$ & $3.8878\times 10^8$    & $7.84\times 10^{-4}$ \\
$^{63}$Ni   &  66.945 & $\beta^-$ & $3.1588\times 10^9$    & $1.38\times 10^{-6}$ \\
$^{93}$Zr   &  60.63  & $\beta^-$ & $4.952\times 10^{13}$  & $2.39\times 10^{-10}$ \\
$^{106}$Ru  &  39.4   & $\beta^-$ & $3.2278\times 10^7$    & $5.88\times 10^{-4}$ \\
$^{107}$Pd  &  33     & $\beta^-$ & $2.0512\times 10^{14}$ & $2.58\times 10^{-10}$ \\
$^{187}$Re  &  2.64   & $\beta^-$ & $1.3727\times 10^{18}$ & $4.32\times 10^{-11}$ \\
\botrule
\end{tabular}
}
\label{cnb:tbl1}
\end{table}

The $\beta$-decay experiments of current generation includes the spectrometer of KATRIN \cite{KATRIN} and the calorimeter of MARE \cite{MARE}.
KATRIN uses 50 $\mu$g of $^{3}$H as the effective target mass, and MARE is planning to deploy 760 grams of $^{187}$Re.
Therefore, we can estimate the C$\nu$B event rates, respectively, as
\begin{eqnarray}
N^{\nu}_{}({\rm
KATRIN})&\simeq&4.2\times10^{-6}\times\sum_{i}|U^{}_{ei}|^2\zeta^{}_i \quad {\rm yr}^{-1}\,,\\
N^{\nu}_{}(\rm
MARE)&\simeq&7.6\times10^{-8}\times\sum_{i}|U^{}_{ei}|^2\zeta^{}_i \quad {\rm yr}^{-1}\,.
\end{eqnarray}
Moreover, a realistic proposal for the C$\nu$B detection is the PTOLEMY project \cite{PTOLEMY}, which is designed to employ 100 grams of $^{3}$H
as the capture target using a combination of a large-area surface-deposition tritium target, the MAC-E filter, the RF tracking,
the time-of-flight systems, and the cryogenic calorimetry. Finally, the event rate of PTOLEMY are calculated to reach the observable level:
\begin{eqnarray}
N^{\nu}({\rm PTOLEMY})\simeq8.0\times\sum_{i}|U^{}_{ei}|^2\zeta^{}_i\quad {\rm
yr}^{-1}\,.
\end{eqnarray}

\section{Flavor Effects}
\label{flavor}

\begin{figure}[t]
\begin{center}
\begin{tabular}{cc}
\includegraphics*[bb=18 18 280 216, width=0.45\textwidth]{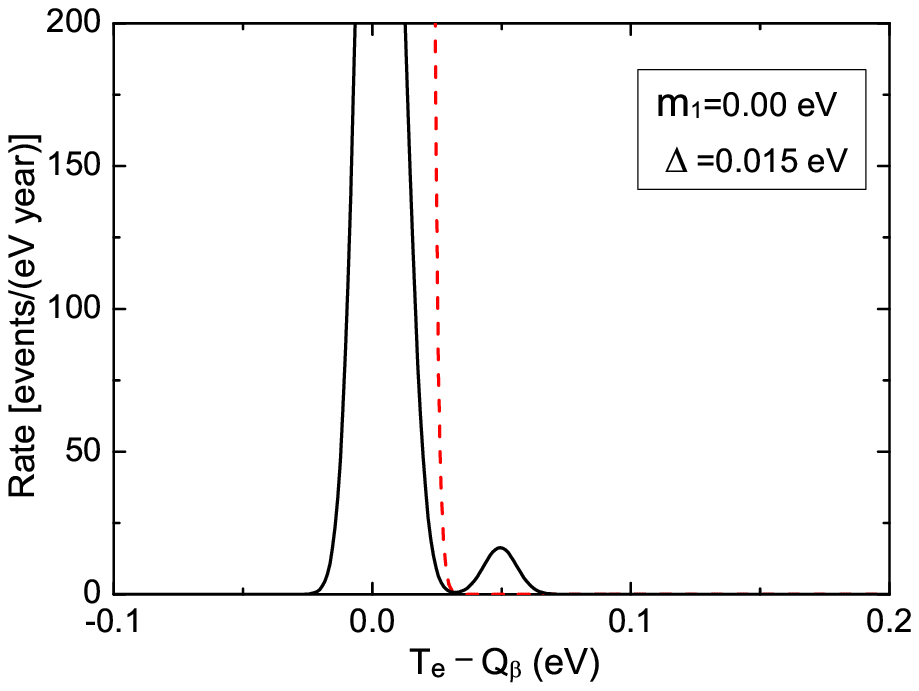}
&
\includegraphics*[bb=18 18 280 216, width=0.45\textwidth]{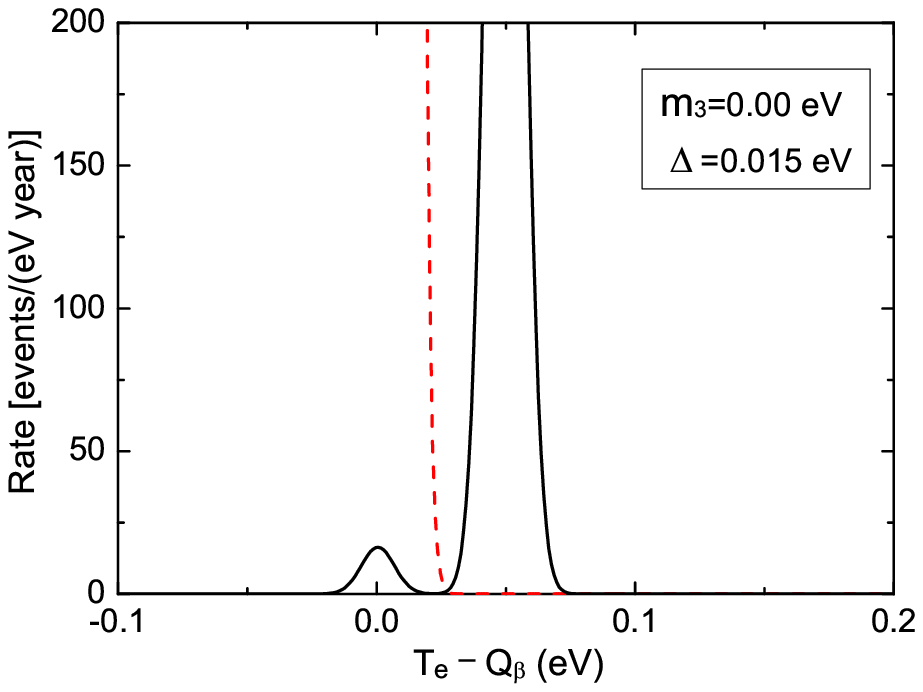}
\\
\includegraphics*[bb=18 18 280 216, width=0.45\textwidth]{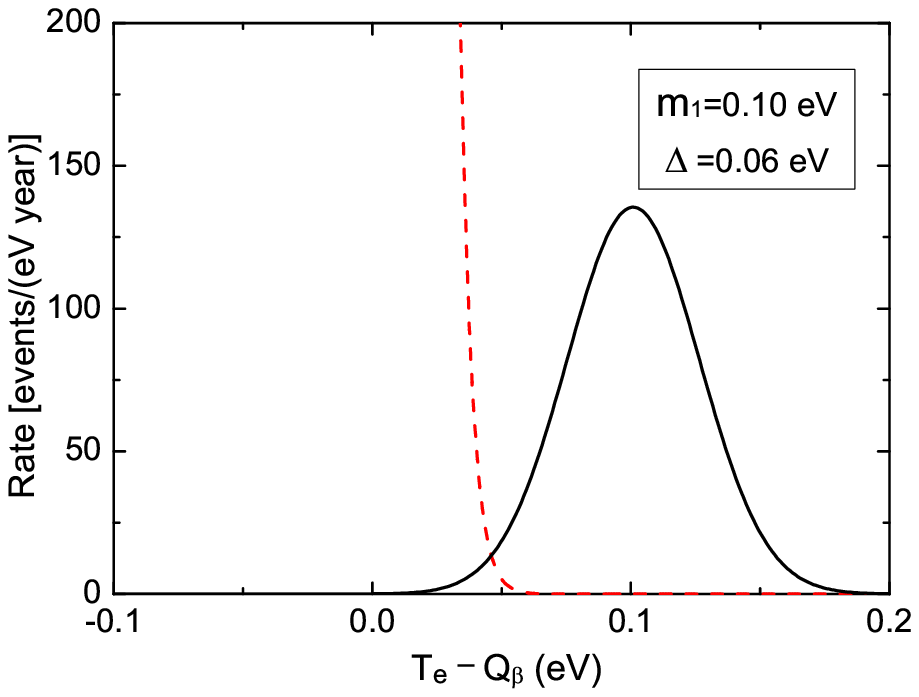}
&
\includegraphics*[bb=18 18 280 216, width=0.45\textwidth]{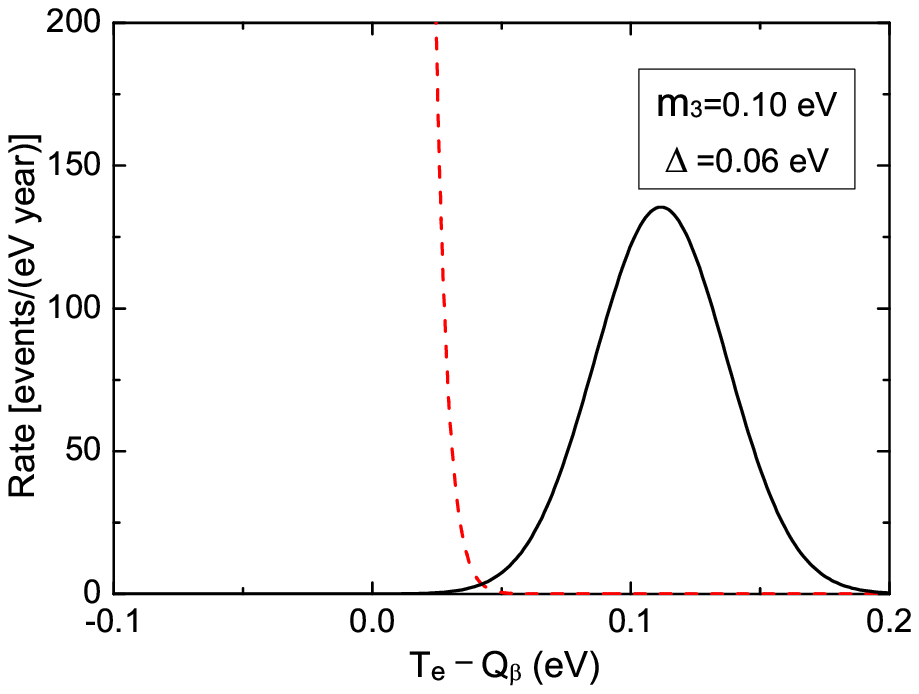}
\end{tabular}
\end{center}
\vspace{-0.5cm}
\caption{The relic neutrino capture rate as a function of the
kinetic energy of electrons in the standard scheme with $\Delta
m^2_{31} > 0$ (left panel) or $\Delta m^2_{31} < 0$ (right panel). The
gravitational clustering of three active neutrinos has been neglected for simplicity.}
\end{figure}
Besides the total capture rates, the C$\nu$B detection exhibits interesting properties of flavor effects
due to the neutrino mixing. In this section, we shall discuss the effects of the neutrino mass hierarchy,
presence of light sterile neutrinos \cite{sterile}, and keV sterile neutrinos as a candidate of warm dark matter \cite{kev}.

In our calculation, we adopt best-fit values of the relevant three-neutrino oscillation parameters
(i.e., $\theta_{12}$, $\theta_{13}$, $\Delta m^2_{21}$ and $|\Delta m^2_{31}|$) from Review of Particle Physics \cite{PDG},
and all the other parameters will be explicitly mentioned when needed. Our default assumption of the target mass is
100 grams of $^{3}$H, but will be 10 kg of $^{3}$H or 1 ton of $^{106}$Ru when we discuss the keV sterile neutrinos.

Fig.~2 shows the capture rate of the C$\nu$B as a function of the kinetic energy $T^{}_e$ of electrons in the standard three-neutrino scheme with
$\Delta m^2_{31} > 0$ (left panel) and $\Delta m^2_{31} < 0$ (right panel).
The finite energy resolution $\Delta$ (i.e., $\Delta = 2\sqrt{2\ln 2} \,\sigma$) is taken in such a way
that only one single peak can be observed beyond the $\beta$-decay background. The
gravitational clustering of three active neutrinos has been neglected for simplicity.
As the lightest neutrino mass ($m^{}_1$ in the left panel or $m^{}_3$ in the right panel)
increases from 0 to 0.1 eV, the neutrino capture signal moves towards the
larger $T^{}_{e}$ region. Hence the distance between the signal peak and the $\beta$-decay background becomes larger for
a larger value of the lightest neutrino mass, and therefore the required energy resolution is less stringent.
Comparing between the left panel and right panel, one can observe that it is easier to detect the C$\nu$B in the
$\Delta m^2_{31} <0$ case, where the capture signal is separated more apparently from the $\beta$-decay background.
The reason is that the dominant mass eigenstates $\nu^{}_1$ and $\nu^{}_2$ in $\nu_{\rm e}$ have greater eigenvalues than in the $\Delta m^2_{31} > 0$ case.

\begin{figure}[t]
\begin{center}
\begin{tabular}{cc}
\includegraphics*[bb=18 18 274 212, width=0.46\textwidth]{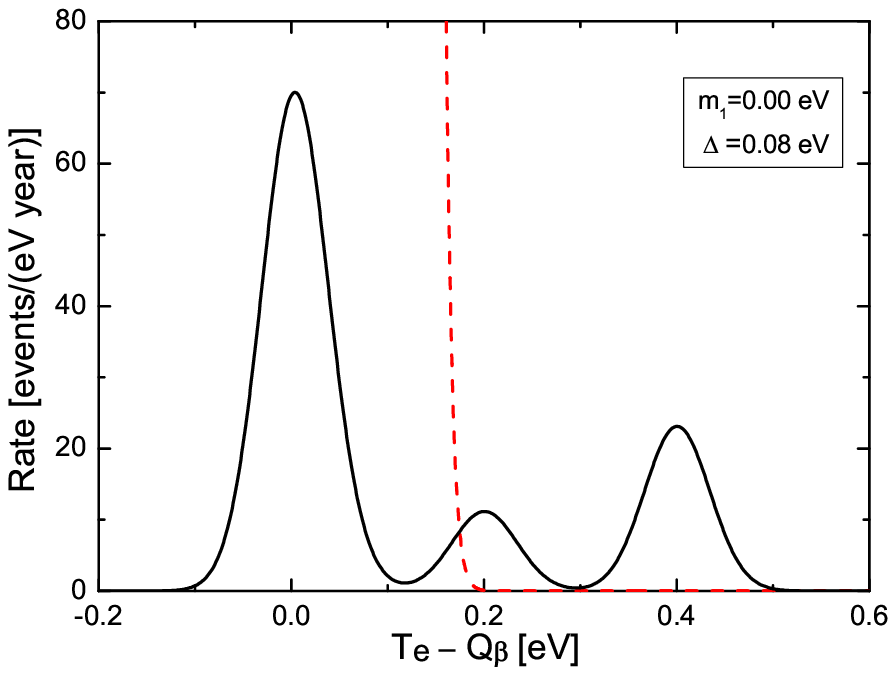}
&
\includegraphics*[bb=18 18 274 212, width=0.46\textwidth]{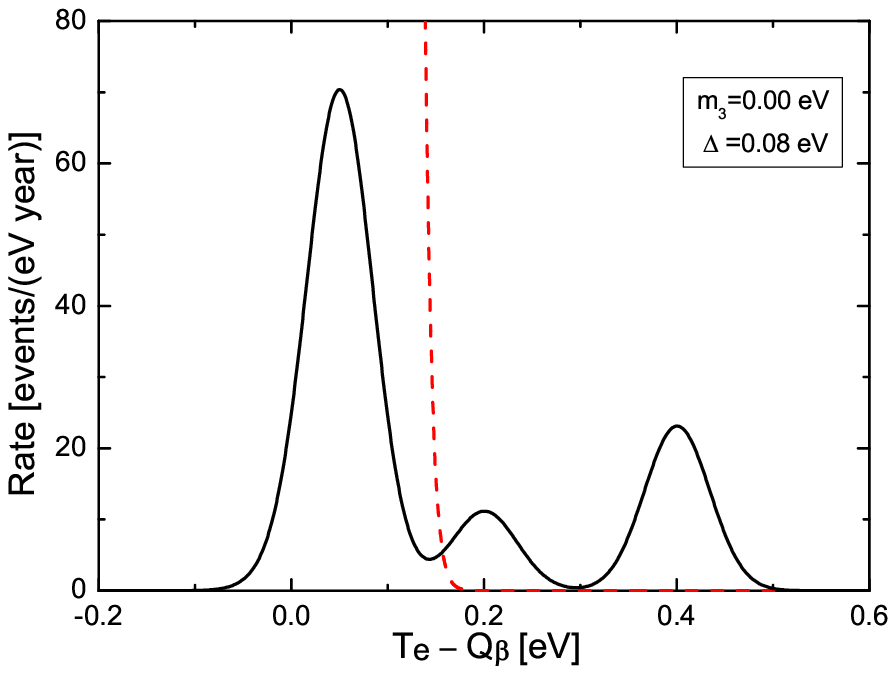}
\end{tabular}
\end{center}
\vspace{-0.5cm}
\caption{The capture rate of the C$\nu$B as a function of the electron's kinetic
energy in the (3+2) mixing scheme with $\Delta m^2_{31} > 0$
(left panel) and $\Delta m^2_{31} < 0$ (right panel) \cite{LLX}. The
gravitational clustering of relic sterile neutrinos around the Earth
has been illustrated by taking $\zeta^{}_1 = \zeta^{}_2 = \zeta^{}_3
=1$ and $\zeta^{}_5 = 2 \zeta^{}_4 = 10$ for example.}
\end{figure}
Next we are going to study the (3+2) mixing scheme with two light sterile
neutrinos. Considering the hints of short baseline oscillations \cite{sterile,sterile2},
we assume $m^{}_4 = 0.2$ eV and $m^{}_5 = 0.4$ eV together with
$|U^{}_{e1}| \approx 0.792$, $|U^{}_{e2}| \approx 0.534$,
$|U^{}_{e3}| \approx 0.168$, $|U^{}_{e4}| \approx 0.171$ and
$|U^{}_{e5}| \approx 0.174$ in the numerical calculations.
We also take $m^{}_1 =0$ or $m^{}_3 =0$ for simplicity.
We illustrate the capture rate of
the C$\nu$B as a function of the electron's kinetic energy $T^{}_e$ against the corresponding $\beta$-decay background for both $\Delta
m^2_{31} > 0$ and $\Delta m^2_{31} < 0$ schemes in Fig.~3. To take account of possible gravitational
clustering effects, we assume $\zeta^{}_1 = \zeta^{}_2 = \zeta^{}_3 =1$ (without clustering
effects for three active neutrinos) and $\zeta^{}_5 = 2\zeta^{}_4 = 10$ (with mild clustering effects for two sterile neutrinos).
As one can see from Fig.~3, the signals of sterile neutrinos are
obviously enhanced because of $\zeta^{}_4 >1$ and $\zeta^{}_5 >1$. If
the overdensity of relic neutrinos is very significant around the Earth,
it will be helpful for the C$\nu$B detection through the neutrino capture process.

\begin{figure}[t]
\begin{center}
\begin{tabular}{cc}
\includegraphics*[bb=18 18 275 216, width=0.46\textwidth]{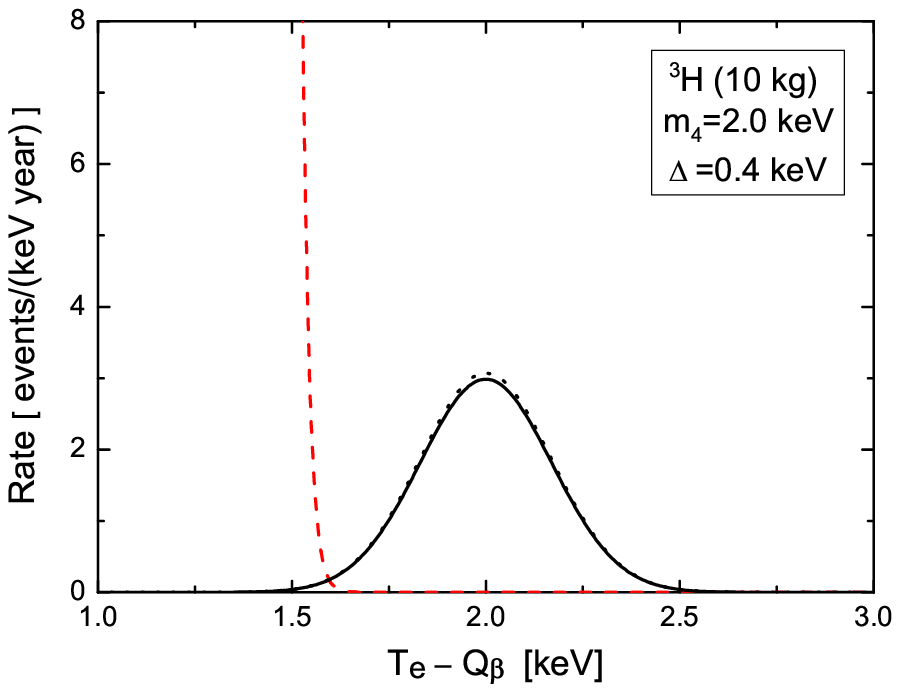}
&
\includegraphics*[bb=18 18 275 216, width=0.46\textwidth]{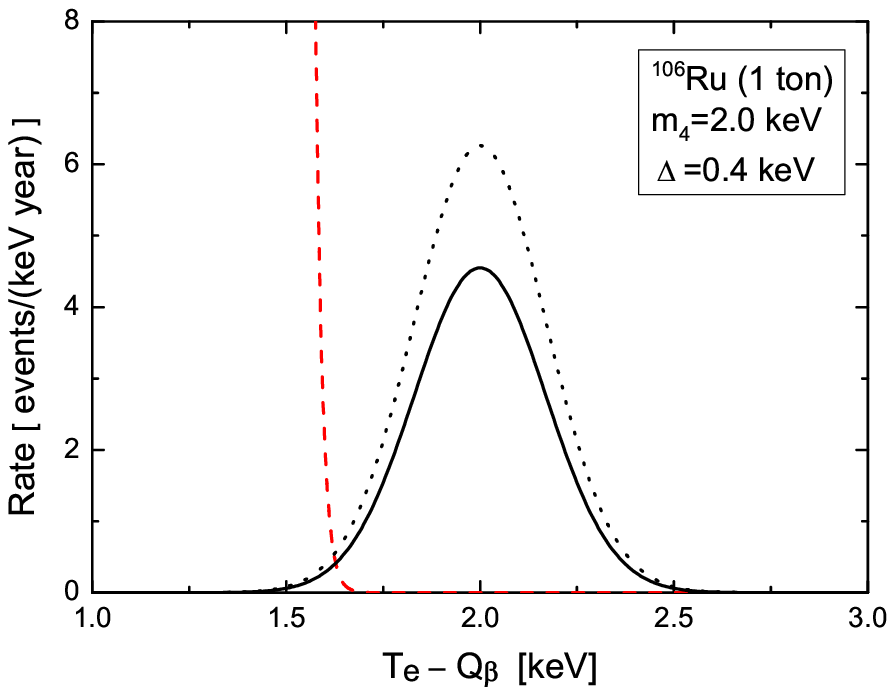}
\end{tabular}
\end{center}
\vspace{-0.5cm}
\caption{The keV sterile neutrino capture rate as a function of the
kinetic energy of electrons with $^3{\rm H}$ (left panel) and
$^{106}{\rm Ru}$ (right panel) as our target sources \cite{LX11}. The solid (or
dotted) curves denote the signals with (or without) the half-life
effect.}
\end{figure}
Finally we want to talk about the detection of keV sterile neutrinos as a candidate of warm dark matter.
We shall work in the (3+1) mixing scheme, where the mixing element and mass of the sterile neutrino are
severely constrained by cosmological observational data \cite{kev}. Here $m^{}_4 = 2 ~{\rm keV}$ and $|U^{}_{e4}|^2 \simeq 5 \times
10^{-7}$ are assumed for illustration. We also take $\Delta m^2_{31} > 0$ and $m^{}_{1}=0$ for simplicity.
From the mass density of dark matter around the Earth \cite{kev}, (i.e., $\rho^{\rm local}_{\rm DM} \simeq 0.3 ~{\rm GeV} \ {\rm
cm}^{-3}$), we can calculate the number density of $\nu^{}_4$ as
\begin{eqnarray}
n^{}_{\nu^{}_4} \simeq 10^{5} \times \frac{3 ~{\rm
keV}}{m^{}_4} ~{\rm cm}^{-3}\,.
\end{eqnarray}
Our numerical calculations are presented in Fig.~3, where two isotope sources (i.e., 10 kg $^3{\rm H}$ and 1 ton $^{106}{\rm Ru}$)
are illustrated for comparison. One can notice that the required energy resolution is not a problem because of the larger sterile neutrino mass.
However, the extremely small active-sterile mixing makes the observability of keV sterile neutrinos rather dim and remote.
We also show the half-life effects of both isotopes in Fig.~4. The finite lifetime is negligible
for the $^3{\rm H}$ nuclei, but important for the $^{106}{\rm Ru}$ nuclei. It can reduce around $30\%$ of
the capture rate on $^{106}{\rm Ru}$. Hence the half-life effect must be considered if duration of the exposure time is comparable
with the source half-life.

\section{Other possibilities}
\label{other}

The capture on the $\beta$-decaying nuclei is one of the most promising methods to detect directly the C$\nu$B. However,
from Eq.~(7) only neutrinos of the electron flavor are relevant for this detection. Therefore, we should consider other possibilities for
relic neutrinos of $\mu$ and $\tau$ flavors and antiparticles of the C$\nu$B.

\begin{figure}[t]
\begin{center}
\begin{tabular}{cc}
\includegraphics*[bb=16 18 278 220, width=0.46\textwidth]{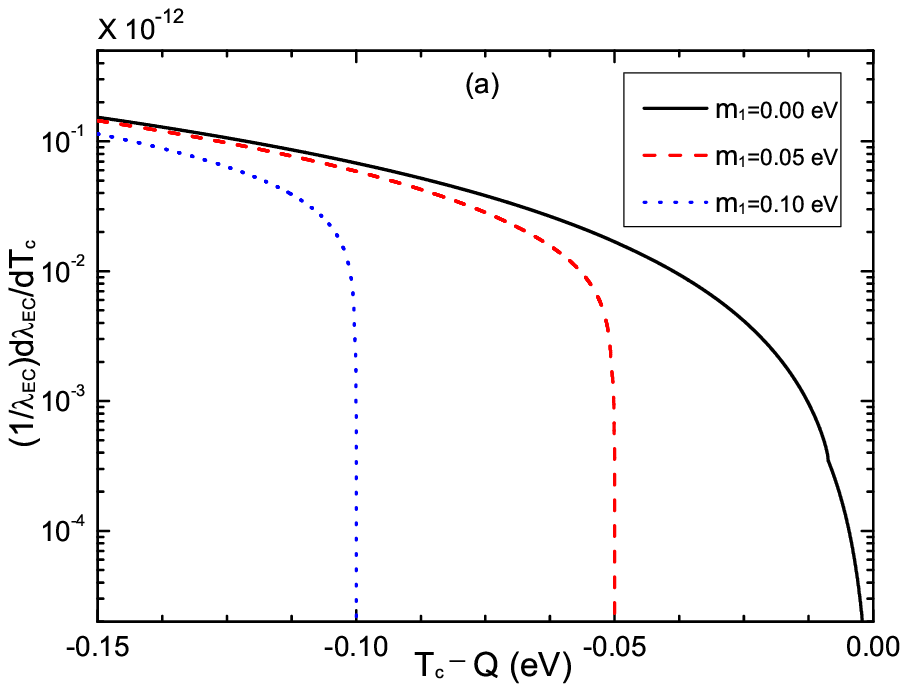}
&
\includegraphics*[bb=16 18 278 220, width=0.46\textwidth]{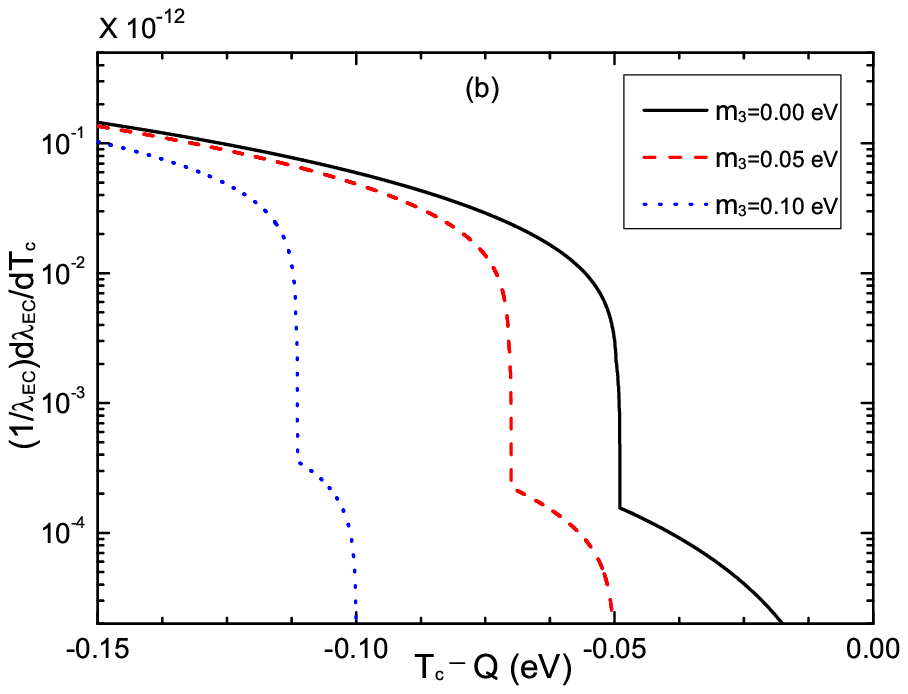}
\\
\end{tabular}
\end{center}
\vspace{-0.5cm}
\caption{The fine structure spectrum near the endpoint of the
$^{163}{\rm Ho}$ EC-decay in the $m^{2}_{31}>0$ (left panel) or
$m^{2}_{31}<0$ (right panel) case \cite{LX11EC}.}
\end{figure}
Similar to the process of captures on $\beta$-decaying nuclei, the EC-decaying nuclei can be
the target of relic antineutrino captures. Here we take the isotope $^{163}{\rm Ho}$ as the
working example \cite{Cocco2,Lusignoli,LX11EC,LXJCAP}.
It should be stressed that the structure near the endpoint of the $^{163}{\rm Ho}$ EC-decay spectrum
is applicable to study the absolute neutrino masses in a similar
way as the $\beta$-decay. We could distinguish between the inverted and normal mass
hierarchies by comparing the right and left panels of Fig.~5. The properties of the relic antineutrino capture
against the EC-decaying background are similar to those discussed in Sec.~2. As the order of magnitude estimate,
one needs 30 kg $^{163}{\rm Ho}$ to obtain one event per year for the relic antineutrino detection, and
needs as much as 600 ton $^{163}{\rm Ho}$ to get one event per year for the keV sterile antineutrino detection.

Another appealing possibility is the annihilation of EHEC$\nu$s with the C$\nu$B
in the vicinity of $Z$-resonance \cite{Weiler82,Eberle04,Barenboim04} (i.e., $\nu \bar\nu\to Z$).
The resonance energy of EHEC$\nu$s associated with each neutrino
mass eigenstate can be calculated as
\begin{eqnarray}
E^{\rm res}_{0,i} =\frac{m^2_{Z}}{2m^{}_{i}} \simeq 4.2\times 10^{12} \left(\frac{1\,{\rm eV}}{m_{\nu_i}}\right)\,{\rm GeV}\,,
\end{eqnarray}
where $m_{Z}$ denotes the $Z$ boson mass. Since the annihilation cross-section at the
resonance is enhanced by several orders of magnitude compared to the non-resonant scattering.
Therefore, the absorption dips at the resonance energies are expected for the EHEC$\nu$s arriving
at the Earth \cite{Weiler82,Eberle04,Barenboim04},
which could provide the direct evidence for the existence of the C$\nu$B.

Recently there is another interesting method of the C$\nu$B detection using
the atomic de-excitation process \cite{atomic}. The de-excitation process of metastable atoms into the emission mode of a single
photon and a neutrino pair, is defined as the radiative emission of neutrino pair (RENP),
\begin{eqnarray}
|e\rangle \to |g\rangle + \gamma + \nu^{}_{i} + \nu^{}_{j}\,,
\end{eqnarray}
where $|e\rangle$ and $|g\rangle$ are the respective initial and final states of the atoms,
$\nu^{}_{i}$ and $\nu^{}_{j}$ are the neutrino mass eigenstates. The existence of the C$\nu$B may distort
the photon energy spectrum because of the Pauli exclusion principle, which provides a possible promising way
to detect the C$\nu$B \cite{atomic}.

\section{Conclusion}
\label{conc}

The standard Big Bang cosmology predicts the existence of a cosmic neutrino background formed
at a temperature of about one MeV and an age of one second after the Big Bang.
A direct measurement of the relic neutrinos would open a new window to the early Universe.
In this review we have discussed the future prospects for the direct detection of the C$\nu$B,
with the emphasis on the method of captures on $\beta$-decaying nuclei and the PTOLEMY project.
We calculated the neutrino capture rate as a function of the
kinetic energy of electrons against the corresponding $\beta$-decay background, and discussed
the possible flavor effects including the neutrino mass hierarchy, light sterile neutrinos,
and keV sterile neutrinos as a candidate of warm dark matter.
Other possibilities using the EC-decaying nuclei, the annihilation
of EHEC$\nu$s at the $Z$-resonance, and the atomic de-excitation method are also presented in this review.
We stress that such direct measurements of the C$\nu$B in the laboratory experiments might not be
hopeless in the long term.

\section*{Acknowledgement}

The author would like to thank the organizers for the kind invitation and warm hospitality in Singapore,
where this wonderful conference was held.
This work was supported by the National Natural Science Foundation of China under grant Nos. 11135009 and 11305193.




\end{document}